
\magnification \magstep1
\input amstex
\documentstyle{amsppt}
\NoRunningHeads
\baselineskip 15pt

\pagewidth{6.4 truein}
\pageheight{8.6 truein}

\define\ve{\varepsilon}

\define\dv{\setbox2=\hbox{{.}\kern-.27em\raise.6ex\hbox{.}
\kern-.615em\raise1.2ex\hbox{.}}%
 \copy2}

\define\ndv{\setbox3=\hbox{\dv\kern-.40em \hbox{/}}%
\copy3}

\define\be{\bold e\,}

\topmatter
\title Does a Quantum Particle Know the Time?
\endtitle
\author Lev Kapitanski and Igor Rodnianski
\endauthor
\affil  Department of Mathematics\\  
Kansas State University\\ 
Manhattan, Kansas 66506 
\endaffil 
\endtopmatter

Consider the Hamiltonian 
$\,H\,=\,-(1/ 4\pi) (\partial^2/\partial x^2)\,$ on the circle 
$\,\Bbb T = \Bbb R/\Bbb Z$. The unitary exponent $\,\exp i t H\,$ 
is the solution operator for the time-dependent Schr\"odinger equation
$$
{1\over i}\,{\partial\hfill\over\partial t}\,E\,+\,
{1\over 4\pi}{\partial^2\hfill\over\partial x^2}\,E\,=\,0, \tag 0.1a
$$
i.e., $\,\exp i t H\,E^0\,(x)\,$ is the solution of (0.1a) 
with  the initial condition $\,E^0$. The distributional kernel 
of $\,\exp i t H\,$ can be written as a series,
$$
\langle x | \exp i t H | y\rangle\,=\,
\sum_{n\in \Bbb Z}\,\be ( {n^2 t\over 2} + n (x-y))\,. 
$$
where we use the notation $\,\be (z)= e^{2 \pi i z}$.    
We replace $\,(x-y)\,$ by $\,x\,$ and look at the function
$$
E(t,x)\,=\,
\sum_{n\in \Bbb Z}\,\be ( {n^2 t\over 2} + n x )\,.\tag 0.2
$$
We will be interested in the regularity of $\,E(t,x)\,$ in $\,x\,$ 
at different times $\,t\in [0,2)$. 

 
It will be convenient to view the functions on $\,\Bbb T\,$ as periodic 
functions on $\,\Bbb R\,$ of period 1. Then (0.2) is the solution of (0.1a) 
corresponding to the initial condition -- a comb, 
$$
E(0,x)\,=\,E^0(x)\,:=\,\sum_{n\in \Bbb Z}\,\delta(x-n).\tag 0.1b
$$
The usual framework for describing the regularity of the solutions of 
the Schr\"odinger equation (0.1a) is the scale 
$\,\{H^s, \;s\in\Bbb R\}\,$ of $\,L^2$-Sobolev spaces. Recall that 
a distribution $\,f(x)=\sum_m f_m\be(m x)\,$ belongs to the space 
$\,H^s\,$ if $\,\sum_m \langle m\rangle^{2s}\,|f_m|^2\,<\infty$, 
where $\,\langle m\rangle = (1+m^2)^{1/2}$. The exponent $\,\exp i t H\,$ 
is a continuous operator in each of the Sobolev spaces $\,H^s$. 
The comb-function (0.1b) lies in $\,H^s$, where $\,s\,$ is any number less 
than $\,-1/2$. For {\it every\/} $\,t>0$, the solution (0.2), 
if viewed through 
the telescope of Sobolev spaces, has the same regularity in $\,x\,$ variable, 
namely, $\,E(t,\cdot)\in\cup_{s<-1/2}\; H^s$. 
\medskip


The Sobolev spaces are not the only function spaces 
that can be applied to the analysis  of the solutions of 
the Schr\"odinger equation. Of special interest for us here 
will be the Besov spaces. We show that the regularity 
of $\,E(t,\cdot)$, when measured in the appropriate 
Besov spaces, changes with $\,t$. The most drastic difference 
in regularity is between the cases when $\,t\,$ is 
rational and when $\,t\,$ is irrational. 
Within the set of irrational times, although there exists a generic 
regularity for generic $\,t$, there are different thin classes 
of irrationals which prescribe their particular regularity 
to the fundamental solution. These classes are singled out 
and characterized by the behavior of the continued fraction 
expansions of their members.
\medskip


Note, that when $\,t\,$ is in the upper 
half-plane, then $\,E(t,x)\,$ defined in (0.2) is 
(essentially) Jacobi's theta-function. The well-known transformation 
properties of theta-functions allow to express 
$\,E(t,x)$, when $\,t$ is rational,  as a linear 
combination of $\,\delta$-functions sitting in 
a (depending on $\,t$) finite number of points $\,x\,$ 
on the circle. This completely answers the question 
of regularity at rational times.


If $\,t\,$ is irrational, the situation is more complicated. 
Our choice of Besov spaces to measure the regularity of 
$\,E(t,\cdot)$, requires the estimates in $\,L^{\infty}\,$ 
of the exponential 
sums of the form 
$$
\sum\limits_{n\in\Bbb Z}\chi(2^{-j}|n|)\,
\be ( {n^2 t\over 2} + n x ),\tag 0.3
$$ 
for large $\,j$. Here $\,\chi\,$ is a cut-off function, which
is supported on the interval $\,[1/2,\,2]$, and which  
is either smooth, or equals  $\,1$ on this interval.
The exponential sums (0.3) have been studied extensively, 
especially during the last 90 years. Of particular importance for 
us are the results of \cite{Hardy \& Littlewood, 1914} with 
subsequent developments of \cite{Mordell, 1926}, and 
\cite{Fiedler, Jurkat \& K\"orner, 1977}, and \cite{Bombieri, 1990}.
\medskip


We should mention that K. I. Oskolkov have obtained some nice 
results on the regularity of certain solutions of the 
Schr\"odinger equation, which exhibit different behavior 
for rational and irrational times, see \cite{Oskolkov, 1992} 
and further references to his works therein. However, the 
questions he discusses and the function spaces he uses 
are different from the ones we concern ourselves in the present 
paper. 

Also related to what we are doing, are the studies of the value distribution 
properites of theta-sums in 
\cite{Jurkat \& van Horn, 1981, 1982}, \cite{Sarnak, 1981, 1982}, 
\cite{Marklof, 1996}, and the studies of the geometric patterns 
generated by theta-sums in \cite{Dekking \& Mend\`es France, 1981}, 
\cite{Deshouillers, 1985}, 
\cite{Berry \& Hannay, 1987}, \cite{Berry \& Goldberg, 1988}, 
\cite{Coutsias \& Kazarinoff, 1987}. 
\medskip


Our interest in the regularity of  $\,E(t,x)\,$ stemmed, 
initially, from 
our previous work on the regularity of the fundamental 
solution for the time-dependent Schr\"odinger equation 
in $\,\Bbb R^n\,$ with a growing at infinity potential. 
There, the model problem is the follwoing:
$$ 
{1\over i}\,{\partial\hfill\over\partial t}\,E\,-\,
\Delta_x E\,+\,|x|^{\rho}\,E\, =\,0,\quad t\in\Bbb R^1,\;x\in\Bbb R^d\,;
\qquad E(0,x,y)\, =\,\delta (x-y)\,,
\tag 0.4 
$$
$d\ge 1$, and $\,\rho\,$ is a positive constant. 
The regularity of $\,E(t,x,y)\,$ depends on the rate of growth of the 
potential. If $\,\rho<2$, then $\,E(t,x,y)\,$ is $\,C^{\infty}$-smooth 
in $\,(t,x,y)$,  $\,t\ne 0$; see \cite{Yajima, 1996}, 
\cite{Kapitanski \& Rodnianski}. In the case $\,\rho=2\,$ the solution, 
$\,E(t,x,y)$, is given by Mehler's formula and shows that the singularities 
reappear at resonant times $\,t=m\pi$, $\,m\in\Bbb Z$, while for 
all other $\,t\,$ the fundamental solution is smooth. This picture survives 
the perturbations of $\,|x|^2\,$ by functions growing slower than 
quadratically, \cite{Kapitanski, Rodnianski, Yajima, 1997}.


When $\,\rho>2\,$ and $\,d>1$, nothing is known about 
the regularity  of $\,E(t,x,y)\,$ (except, of course, for what the standard 
energy estimates give). However, it is likely that 
the fundamental solution is nowhere smooth. This conjecture is 
supported by a remarkable recent result of K.~Yajima, who showed 
that in the one-dimensional case ($d=1$), if $\,\rho>2$, then 
$\,E(t,x,y)\,$ is not even in the local Bessel-Sobolev space 
$\,\Cal L^{1/\rho}_{1,\,loc}(\Bbb R^3)$, see \cite{Yajima, 1996}. 


The initial boundary value problem (IBVP) for the Schr\"odinger equation 
may be viewed as the extreme limit case of (0.4) as $\,\rho\to\infty$.
Yajima's technique, when applied to the one-dimensional IBVPs, 
gives the corresponding nonsmoothness results. In particular, 
it turns out that
the distributional kernel $\,E(t,x,y)\,$ 
of the operator $\,\exp\{-i t \Cal H\}$, where 
$\,\Cal H=-(1/4\pi)\,d^2/dx^2\,$ on the interval $\,[0,1/2]\,$ 
with  Dirichlet boundary conditions, i.e., the function 
$\,E(t,x,y)=4\sum_{n=1}^{\infty} 
e^{-\pi itn^2}\sin\,2\pi nx\,\sin\,2\pi ny\,$ 
is nowhere locally integrable (in $\,\Bbb R^3$), 
see \cite{Yajima}, Remark 4. 
Our results will add to this by revealing the fine changes 
in regularity of $\,E(t,\cdot,\cdot)\,$ at different times $\,t$. 
\medskip


{\bf ACKNOWLEDGMENTS.\/} This work was partially supported by NSF grant DMS-9623520.  
The authors thank Peter Sarnak and Jens Marklof 
for valuable discussions.
The second author also thanks the Institute for Mathematics and 
Its Applications at the University of Minnesota, and the organizers of 
the workshop ``Emerging Applications of Number Theory" at IMA, for their 
hospitality.

\medskip


\head {\bf 1\/}\ \  Statements of  Results \endhead

We start with the definitions of the Besov spaces 
we need to state the results. For reference on Besov spaces 
see \cite{Bergh \& L\"ofstr\"om}, \cite{Triebel}.

Let $\, \chi\,$ be a $\,C_0^{\infty}\,$ 
function on $\,\Bbb R_+\,$ with the following properties: 
$$
\text{supp}\,\chi\,=\,[2^{-1}, 2], \tag 1.1a
$$
and
$$
\sum\limits_{j=-\infty}^{\infty}\,\chi (2^{-j}\xi)\,=\,1,
\qquad\forall \xi >0.\tag 1.1b
$$
Define the functions 
$$
\chi_j(\xi) =\chi (2^{-j}\xi), \qquad j=1,\,2, \dots;\qquad 
\chi_0(\xi) = 
1-\sum\limits_{j=0}^{\infty}\,\chi_j (\xi)\,. \tag 1.2
$$
With each of these functions we associate an operator $\,K_j$, which 
maps a distribution $\,f(x)=\sum_m f_m\, \be ( m x)\,$
to a (finite) exponential sum,
$$
K_j(f)(x)= 
\sum_{m=-\infty}^{\infty}\, \chi_j(|m|)\, f_m\, \be ( m x),
\quad j=0,\,1,\dots.\tag 1.3a
$$

We define the Besov spaces $\,B^s_{u,v}\,$ on $\,\Bbb T$ 
for the following values of parameters $\,s$, $\,u$, and $\,v$:\ 
$\;-\infty<s<+\infty$, 
$\,1\le u,\,v \le +\infty$. By definition,  
the space $\,B^s_{u,v}\,$ is composed of 
all distributions $\,f(x)=\sum_m f_m\, \be ( m x)\,$ 
such that the following norm is finite:
$$
\align
\| f\|_{{}_{ B^s_{u,v}}}\,&=\,
\bigg(\,\sum_{j=0}^{\infty}2^{s j v}\,
\|\, K_j(f)(\cdot)\, \|_{{}_{L^u}}^v \bigg)^{\frac 1v},
\qquad \text{if}\quad v<\infty,\tag 1.4a \\
\| f\|_{{}_{ B^s_{u,\infty}}}\,& =\,\sup_{j\ge 0}\;
 2^{s j }\,
\|\,  K_j(f)(\cdot)\, \|_{{}_{L^u}},
\qquad \quad\quad\;\;\text{if}\quad v=\infty.\tag 1.4b
\endalign 
$$

Along with the standard Besov spaces $\,B^s_{u,v}$, we 
use what we call the rough Besov spaces $\,\lceil B\rceil^s_{u,v}$. 
Their definition differs from that of $\,B^s_{u,v}$, 
practically, only 
in the choice of the cut-off function: instead of 
a smooth $\,\chi$, 
we take a step-function $\,\lceil\chi\rceil$ -- 
the characteristic function 
of the interval $\,[2^{-1}, 2]$. To be more precise, define the 
operators $\,\lceil K\rceil_j\,$ for $\,j=0,\,1,\dots $, as follows:
$$\split 
\lceil K\rceil_0(f)(x)=\sum_{m=-2}^2 f_m\, \be ( m x);
\\ 
\lceil K\rceil_j(f)(x)=\sum_{2^{j-1}< |m|\le 2^{j+1}} f_m\, \be ( m x)\,,
\quad j>0.
\endsplit
\tag 1.3b
$$
The norm in $\,\lceil B\rceil^s_{u,v}$ is defined as in 
(1.4), with $\,K_j\,$ replaced by $\,\lceil K\rceil_j$. 
It is known that if $\,1<u<\infty$, then the spaces 
$\, B^s_{u,v}\,$ and $\,\lceil B\rceil^s_{u,v}\,$ 
consist of the same distributions, and the norms are equivalent 
(see \cite{Lizorkin, 1977}, \cite{Triebel, 2.5.4 and 9.1.3}). 
Recall, also, that $\,B^s_{2,2}=\lceil B\rceil^s_{2,2}=H^s(\Bbb T)$.


In this paper we use the spaces 
$\,B^s_{\infty,\infty}\,$ and  $\,\lceil B\rceil^s_{\infty,\infty}\,$ 
most of all. Therefore, we make a simpler notation for 
them: $\,B^s_{\infty}\,$ 
and $\,\lceil B\rceil^s_{\infty}$.


We shall say that a distribution $\,f\,$ {\it belongs to $\,B^s_{\infty}\,$ 
sharp\/} if $\,f\in B^s_{\infty}$ and 
$\,f\notin \cup_{\epsilon>0}\,B^{s+\epsilon}_{\infty}$. 
A similar convention will be in force for the spaces 
$\,\lceil B\rceil^s_{\infty}$ and $\,H^s$. 


Closely related to the persistence of regularity 
of solutions of (0.1a) in $\,H^s\,$ is the following fact. 

\proclaim{Remark 1.1} For all $\,t$, 
the distribution $\,E(t,\cdot)\,$ belongs to 
$\,B^{-1/2}_{2,\infty}\,$ sharp.
\endproclaim

\proclaim{Remark 1.2}
That $\,E(t,\cdot)\,$ belongs to the Besov space 
$\,\lceil B\rceil^{-\alpha}_{\infty}\,$ is equivalent to the existence of a 
constant $\,\lceil C\rceil\,$ such that
$$
\sup_x\,
\big | \sum\limits_{n=2^{j-1}}^{2^{j+1}} \be ({n^2t\over 2}+nx)\big |
\le \lceil C\rceil\,2^{\alpha j},
\quad\text{\rm for all sufficiently large}\quad j\,.
\tag 1.5a
$$

That $\,E(t,\cdot)\,$ belongs to the Besov space 
$\, B^{-\alpha}_{\infty}\,$ is equivalent to the existence of a 
constant $\,C\,$ such that
$$\multline
\sup_x\,
\big | \sum\limits_{n=2^{j-1}}^{2^{j+1}}
\chi(2^{-j}n) \be ({n^2t\over 2}+nx)+
\chi(2^{-j}n) \be ({n^2t\over 2}-nx)
\big |
\le C\,2^{\alpha j},\\
\hfill\text{\rm for all sufficiently large}\quad j\,.\hfill
\endmultline
\tag 1.5b
$$
\endproclaim
\qed


The regularity properties of $\,E(t,\cdot)\,$ in Besov spaces 
$\,B^s_{\infty}\,$ and $\,\lceil B\rceil^s_{\infty}$ 
depend on the 
continued fraction representation of $\,t$. 
We refer the reader to \cite{Khinchin, 1964} and \cite{Schmidt, 1980} 
for the basic 
theory of continued fractions. 
Consider first the case of a rational $\,t=\frac pq \in [0,2)$. 
In this case $\,t\,$ has a finite continued fraction expansion:
$$
t=\frac pq=[a_0,a_1,\dots , a_n]\,=\,a_0+\,
{1\over\displaystyle a_1+
{\strut 1\over\displaystyle \dots
{\strut \dots\over
{\strut 1\over\displaystyle a_n}}}}\;. 
$$
Note, that the expansion is not unique: we also have 
$\,\frac pq = [a_0,a_1,\dots, a_{n-1}, a_n-1,1]$, if $\,a_n\ne 1$, 
and $\,\frac pq = [a_0,a_1,\dots, a_{n-1}+1]$, if $\,a_n=1$. 


If $\,t=[a_0,a_1,\dots , a_n]$, then the numbers 
$\,{p_k/ q_k}\,=\,[a_0,a_1,\dots , a_k]$, 
$\,k=1,\dots, n-1$, are the corresponding convergents, and 
$\,p_n=p$, $\,q_n=q$. 

\proclaim{Theorem {\rom I\/}} Let $\,t\in [0,2)$ be a rational number, 
and $\,t=\frac pq$ -- its simple fraction representation. 
Let $\,{p_k\over q_k}$, $\,k=1,\dots, n-1$, be partial convergents 
to $\,t$ determined by a finite continued fraction 
$\,[a_0,a_1,\dots , a_n]$ with an odd number of quotients 
(i.e., $\,n\,$ is even). Then, 

1) (formula)
$$
E(\frac pq,x)={\varkappa_0(t)\over\sqrt{q}}\,
\be (-\frac 12 q_{n-1}q x^2+
\frac 12 q\eta x -
\frac 18 \xi\eta)\,
\sum\limits_{n} \delta ({n+\frac 12\xi\over q} - x)\,, \tag 1.6
$$
where 
$$
\xi = p\cdot q\;(\text{\rm mod\/}\;2),
\qquad \eta = p_{n-1}\cdot q_{n-1}\;(\text{\rm mod}\;2)\,,
$$
and $\,\varkappa_0(t)\,$ is an eighth root of $\,1$;

2) (regularity)
$$
E(t,\cdot)\,\in\,B^{-1}_{\infty}\cap \lceil B\rceil^{-1}_{\infty}
\quad
\text{sharp}\,.\tag 1.7
$$ 
\endproclaim
\medskip

When $\,t\in(0,2)\,$ is irrational, its continued fraction expansion 
$\,t=[a_0,a_1,\dots , a_n,\dots]\,$ 
is infinite and unique.
 
The regularity of $\,E(t,\cdot)\,$ for generic irrational $\,t\,$ 
is given by the following 
\proclaim{Theorem {\rom {II}}} 

(i) For almost all irrational $\,t$, 
$\;
E(t,\cdot)\,\in\,\cap_{\ve>0} ( B^{-1/2-\ve}_{\infty}\cap \lceil B\rceil^{-1/2-\ve}_{\infty} )\,$. 

(ii) If $\,t\,$ is an irrational number with bounded quotients, 
i.e., 
there is a constant $\,C>0\,$ such that $\,a_n\le C$, for all $\,n$, 
then $\;
E(t,\cdot)\,\in\,B^{-1/2}_{\infty}\cap \lceil B\rceil^{-1/2}_{\infty}
\quad
\text{sharp}$.

(iii) There is no $\,t\,$ for which 
$\,E(t,\cdot)\,$ belongs to 
$\, B^{(-1/2) +\ve}_{\infty}\cup \lceil B\rceil^{(-1/2)+\ve}_{\infty}$ with 
any positive $\,\ve$.
\endproclaim
\medskip

We now define a few  (narrower) classes of irrational numbers
using restrictions on the growth of the denominators $\,q_n$ of 
their convergents $\,p_n/q_n$. 

For $\,\sigma\ge 0$, denote by $\,\Cal I(\le\sigma)\,$ the set of all 
irrational $\,t\,$ such that for each of them there exists a constant 
$\,C_t\,$ such that 
$$
q_{n+1}\,\le\,C_t\,q_n^{1+\sigma}\,,\qquad
\text{for all sufficiently large $\,n$}.\tag 1.8 
$$
Denote by $\,\Cal I(\ge\sigma)\,$ the set of all 
irrational $\,t\,$ such that for each of them there exists a constant 
$\,c_t>0\,$ such that 
$$
q_{n+1}\,\ge\,c_t\,q_n^{1+\sigma}\,,\qquad
\text{for an infinite number of $\,n$}.\tag 1.9
$$
Finally, denote 
$\,\Cal I(\sigma)\,=\,\Cal I(\le\sigma)\,\cap\,\Cal I(\ge\sigma)$.

\proclaim{Theorem {\rom {III}}} 

(i) If $\,t\in\Cal I(\le\sigma)\,$, then 
$$
E(t,\cdot)\in B^{{}^{-{1+\sigma\over 2+\sigma}}}_{\infty}\,\cap\,
\lceil B\rceil^{{}^{-{1+\sigma\over 2+\sigma}}}_{\infty}\,. \tag 1.10a
$$

(ii) If $\,t\in\Cal I(\ge\sigma)\,$, then 
$$
E(t,\cdot)\notin 
\left(
\cup_{\epsilon>0}B^{{}^{-{1+\sigma\over 2+\sigma}+\epsilon}}_{\infty}
\right)\,
\cup\,
\left(\cup_{\epsilon>0}\lceil 
B\rceil^{{}^{-{1+\sigma\over 2+\sigma}+\epsilon}}_{\infty}
\right)\,. \tag 1.10b
$$

(iii) If $\,t\in\Cal I(\sigma)\,$, then 
$$
E(t,\cdot)\in B^{{}^{-{1+\sigma\over 2+\sigma}}}_{\infty}\,\cap\,
\lceil B\rceil^{{}^{-{1+\sigma\over 2+\sigma}}}_{\infty}\quad
\text{sharp}. \tag 1.10c
$$
\endproclaim
\medskip


\head {\bf 2\/}\ \ Estimates for Exponential Sums \endhead
\medskip

In this section we study two basic exponential sums, 
$$
\lceil S\rceil_M^N\,(t,x)\,=\,
\sum_{M\le |n|\le N}\,\be\,({n^2t\over 2}\,+\,n x)\,,\tag 2.1
$$
and 
$$
 S_M^N\,(t,x)\,=\,
\sum_{M\le |n|\le N}\,\omega_n\,\be\,({n^2t\over 2}\,+\,n x)\,.\tag 2.2
$$
The first sum is needed for the estimates of the norm of $\,E(t,\cdot)\,$ 
in the space $\,\lceil B\rceil^s_{\infty}$. The second 
is for the space $\, B^s_{\infty}$. There, the r\^ole of coefficients $\,\omega_n\,$ 
will be assigned to $\,\chi(2^{-j}|n|)$, see (0.3). 
However, in this section we do not restrict ourselves 
to this particular choice of $\,\omega_n$. 


The estimates from above on 
$\,|\sum_{n=1}^N\,\be\,({n^2t\over 2}\,+\,n x)|\,$ 
go back to \cite{Hardy \& Littlewood}. They introduced the method 
of an approximate functional equation for incomplete theta-sums. 
This method was developed further by \cite{Fiedler, Jurkat \& K\"orner}, 
who established, in particular, the result we need, Theorem 2.1 below. 
This result was later proved by \cite{Bombieri} 
using a different technique 
(of maximal operators and Hunt-Carleson theorem).

\proclaim{Theorem 2.1} Let $\,t\,$ be real, and 
$\,|t-\frac pq|\le {1\over q^2}$, for some  co-prime integers 
$\,p\,$ and $\,q$. Then there exists a constant $\,C>0\,$ such that,
for all real $\,x$, 
$$
|\sum_{n=1}^N\,\be\,({n^2t\over 2}\,+\, n x)|\,\le\,
C\,\left( {N\over \sqrt{q}}\,+\,\sqrt{q}\right)\,,\tag 2.3
$$
for any integer $\,N>0$.
\endproclaim

This theorem implies immediately the following estimate.

\proclaim{Corollary 2.2} In the assumptions of Theorem 2.1, 
$$
\| \lceil S\rceil_M^N\,(t,\cdot)\|_{L^{\infty}}\,\le\,
\lceil C\rceil\,\left( {N-M\over\sqrt q}\,+\,\sqrt{q}\right)\,,\tag 2.4
$$
for all integers $\,M\,$ and $\,N>M$.
\endproclaim
\medskip


The estimate from above on the sum (2.2) is given by the following theorem 
(compare \cite{Bourgain, Lemma 3.18}).

\proclaim{Theorem 2.3}  Let $\,t\,$ be real, and 
$\,|t-\frac pq|\le {1\over q^2}$, for some  co-prime integers 
$\,p\,$ and $\,q$. 

Assume that the coefficients $\,\omega_n\,$ 
satisfy the following conditions:
$$
\omega_n = 0,\quad\text{for}\; n\,<M\;\text{and}\; n\,>N,\tag 2.5a
$$
and
$$
\sum_M^N\,|\omega_{n+1}-\omega_n|\,\le\,\varkappa\,. \tag 2.5b
$$
Then   
$$
\| \sum_M^N\,\omega_n\,\be\,({n^2t\over 2}+n\cdot)\|_{L^{\infty}}\,\le\,
\varkappa\,
\lceil C\rceil\,\left( {N-M\over\sqrt q}\,+\,\sqrt{q}\right)\,,\tag 2.6
$$
where $\,\lceil C\rceil\,$ is the same as in (2.4).
\endproclaim
\demo{Proof} Using summation by parts, we obtain
$$\multline 
|\,\sum_M^N\,\omega_n\,\be\,({n^2t\over 2}+n x)\,|\,=\,
|\,\sum_M^N\,(\omega_{n+1}-\omega_n)\,
\sum_{k=M}^n\,\be\,({k^2t\over 2}+k x)\,|\,\\ 
\le\,
\varkappa\,\sup_{M\le n\le N}\,
|\,\sum_{k=M}^n\,\be\,({n^2t\over 2}+k x)\,|,\hfill
\endmultline
$$
and (2.6) follows from (2.4).
\qed
\enddemo

\proclaim{Corollary 2.4}  Let $\,t\,$ be real, and 
$\,|t-\frac pq|\le {1\over q^2}$, for some  co-prime integers 
$\,p\,$ and $\,q$. 

Assume that the coefficients $\,\omega_n\,$ 
satisfy the following conditions:
$$
\omega_n = 0,\quad\text{for}\; |n|<M\;\text{and}\; |n|>N,\tag 2.5a
$$
and
$$
\sum_{M\le |n|\le N}\,|\omega_{n+1}-\omega_n|\,\le\,\varkappa\,. \tag 2.5b
$$
Then 
$$
\| S_M^N(t,\cdot)\|_{L^{\infty}}\le \,
 2\,\varkappa\,
\lceil C\rceil\,\left( {N-M\over\sqrt q}\,+\,\sqrt{q}\right)\,.\tag 2.7
$$
\endproclaim 
\medskip


For the sharpness part in Theorems {\rom I - III\/} we need to 
estimate the exponential sums from below. We do this in two steps.
First, we estimate $\lceil S\rceil_M^N\,(t,x)\,$ and 
$\, S_M^N\,(t,x)\,$ for {\it rational\/} $\,t$, and then show 
that the (suprema of the) sums for an irrational $\,t\,$ are 
close to those with a sufficiently good rational approximation to 
$\,t$. This approach is quite standard, see \cite{Montgomery, 
Chapter 3}. 
\medskip


\proclaim{Theorem 2.5} Let $\lceil S\rceil_M^N\,(t,x)\,$ 
and $\, S_M^N\,(t,x)\,$ be as in (2.1), (2.2). Assume that 
$\,\omega_n$ satisfies (2.5a). 
Let $\,p\,$ and $\,q\,$ be co-prime positive integers. 

For each of four cases below, there exists $\,x=\frac hq\in [0,1]$, 
with integer $\,h$,  
such that
$$
\align
| S_M^N\,(\frac pq, x) |\,& \ge\,
{ \sum\limits_{n=M}^N \omega_n+\omega_{-n} \over \sqrt 2\sqrt{N-M} }\,,
\qquad
\text{if}\quad
q\ge N-M+1\,; \tag 2.8a\\
| S_M^N\,(\frac pq, x) |\,& \ge\,
{ \sum\limits_{n=M}^N \omega_n+\omega_{-n} \over \sqrt{2q} }\,,\qquad
\text{if}\quad
q< N-M+1\,;\tag 2.8b \\
| \lceil S\rceil_M^N\,(\frac pq, x) |\,& \ge\,
\sqrt{2}\sqrt{N-M}\,,\;\qquad\qquad
\text{if}\quad
q\ge N-M+1\,;\tag 2.9a \\
| \lceil S\rceil_M^N\,(\frac pq, x) |\,& \ge\,
{\sqrt{2}(N-M)\over \sqrt{q}}\,,\qquad\qquad
\text{if}\quad
q < N-M+1\,.\tag 2.9b
\endalign
$$
\endproclaim

\demo{Proof} Because of the orthogonality relations among the additive 
characters $(\text{\rm mod}\,2q)$, we have the following equalities:
$$
\sum\limits_{h=1-(M-1)p}^{2q-(M-1)p}\,
\left|S_M^N(\frac pq,\frac hq)\right|^2\, =\,
2q\,\sum\limits_M^N\,|\omega_n|^2+|\omega_{-n}|^2\,,
\tag 2.10a
$$
if $\,q\ge N-M+1$, 
and
$$ 
\sum\limits_{h=1-(M-1)p}^{2q-(M-1)p}\,
\left| S_M^N(\frac pq,\frac hq) \right|^2\,=\,
2q\,\sum\limits_{k=1}^{2q}\,
\left| \sum\limits_{m=0}^{[{N-M+1\over 2q}]}\,
\omega_{2mq+k+M-1}\,+\,
\omega_{-2mq-k-M+1} \right|^2\,,   
\tag 2.10b
$$
if $\,q < N-M+1$.
Since 
$$
|\sum\limits_M^N \omega_n+\omega_{-n}\,|^2\le 
2 (N-M)\,\sum\limits_M^N |\omega_n|^2+|\omega_{-n}|^2,
$$
(2.10a) yields
$$
\sum\limits_{h=1-(M-1)p}^{2q-(M-1)p}\,
\left|S_M^N(\frac pq,\frac hq)\right|^2\,\ge\,
q\,
{|\sum\limits_M^N \omega_n+\omega_{-n}\,|^2
\over N-M}\,.
$$
Hence, 
$$
\left|S_M^N(\frac pq,\frac hq)\right|^2\,\ge\,
{|\sum\limits_M^N \omega_n+\omega_{-n}\,|^2
\over 2(N-M)},
$$
for at least one $\,h$, and (2.8a) follows. 


Similarly, the inequality
$$\multline 
|\sum\limits_M^N \omega_n+\omega_{-n}\,|^2= 
\left| \sum\limits_{k=1}^{2q}
\sum\limits_{m=0}^{[{N-M+1\over 2q}]}\,
\omega_{2mq+k+M-1}\,+\,
\omega_{-2mq-k-M+1} \right|^2\,\le \\
2q\,\sum\limits_{k=1}^{2q}\,
\left| \sum\limits_{m=0}^{[{N-M+1\over 2q}]}\,
\omega_{2mq+k+M-1}\,+\,
\omega_{-2mq-k-M+1} \right|^2\,,
\endmultline
$$
and  (2.10b), show that for at least one integer $\,h$ 
we have (2.8b) with $\,x=\frac hq$.

 
The estimates (2.9) follow from (2.8) if we choose 
$\,\omega_n\,$ to be $\,1\,$ for $\,n\in [-N,-M]\cup [M,N]$, 
and $\,0\,$ otherwise. 
\qed
\enddemo
\medskip

\proclaim{Theorem 2.6} Assume that $\,t\,$ and $\,t_1\,$ are such that 
$$
|t-t_1|< {K\over N^2}\,,\tag 2.11
$$
with some constant $\,K>0$. Then 
$$
\lceil c\rceil_1\,\max_x\,|\,\lceil S\rceil_M^N(t_1,x)\,|\,
\le\,\max_x\,|\,\lceil S\rceil_M^N(t,x)\,|\,\le\,
\lceil c\rceil_2\,\max_x\,|\,\lceil S\rceil_M^N(t_1,x)\,|\,,\tag 2.12a
$$
and
$$
 c_1\,\max_x\,|\, S_M^N (t_1,x)\,|\,
\le\,\max_x\,|\, S_M^N (t,x)\,|\,\le\,
 c_2\,\max_x\,|\,S_M^N (t_1,x)\,|\,,\tag 2.12b
$$
for some constants 
$\,c_1$, $\,c_2$, $\,\lceil c\rceil_1$, and $\,\lceil c\rceil_2$, 
which depend only on $\,K$. 
\endproclaim
\demo{Proof} The proof follows the same line as the proof of Theorem 3.3 
of \cite{Montgomery}. 
First, we define an auxiliary, period 2 function  $\,f(\theta)$:
$$
f(\theta)=\left\{
\aligned
\,&\be\big({t-t_1\over 2}\,(2N\theta)^2\big)\,,
\quad\text{for}\quad 0\le\theta\le\frac 12,\\
\,&\be\big({t-t_1\over 2}\,N^2\big)
\qquad\quad\text{for}\quad \frac 12\le\theta\le 1\,,
\endaligned
\right.
$$
and $\,f(\theta)=f(-\theta)$, for $\,-1\le\theta\le 0$. Taking into 
account (2.11), it is easy to 
check that the total variation of the first derivative of 
$\,f\,$ is bounded, $\,\text{Var}\,f^{\prime}\le \gamma_1$, 
and the Fourier coefficients of $\,f\,$ decay as $\,m^{-2}$, 
$\,|\hat f(m)|\le \gamma_2/m^2$. Note, that 
the constants $\,\gamma_1\,$ and $\,\gamma_2\,$ do not depend on 
$\,t$, $\,t_1$, and $\,N$. 


Choose an $\,x\,$ so that 
$\,|S_M^N(t,x)|\,=\,\max_y\,|S_M^N(t,y)|$. Then we have
$$\multline
|S_M^N(t,x)|=|\sum_{n=M}^{N}\,\omega_n\,\be\,({n^2t\over 2}\,+\,n x)\,
+ \omega_{-n}\,\be\,({n^2t\over 2}\,-\,n x)|=\\
|\sum_{n=M}^{N}\,\omega_n\,f({n\over 2N})\be\,({n^2t_1\over 2}\,+\,n x)\,
+ \omega_{-n}\,f({n\over 2N})\be\,({n^2t_1\over 2}\,-\,n x)|=\\
|\sum_m\,\hat f(m)\,\big(
\sum_{n=M}^{N}\,\omega_n\,\be\,({n^2t_1\over 2}\,+\,n (x+{mn\over 2N}))\,
+ \omega_{-n}\,\be\,({n^2t_1\over 2}\,-\,n (x-{mn\over 2N}))|=\\ 
(\;\text{note, that}\; \hat f(m)=\hat f(-m)\;)\\
|\sum_m\,\hat f(m)\,\big(
\sum_{n=M}^{N}\,\omega_n\,\be\,({n^2t_1\over 2}\,+\,n (x+{mn\over 2N}))\,
+ \omega_{-n}\,\be\,({n^2t_1\over 2}\,-\,n (x+{mn\over 2N}))|=\\
|\sum_m \hat f(m)\,S_M^N(t_1,{x+mn\over 2N})|\le 
\max_y\,|S_M^N(t_1,y)|\cdot \sum_m |\hat f(m)|\le\,
c_2\,\max_y\,|S_M^N(t_1,y)|\,.
\endmultline
$$
The opposite inequality follows by reversing the r\^oles of $\,t\,$ 
and $\,t_1$. Also, (2.12a) is a particular case of (2.12b).
\qed
\enddemo



\head {\bf 3\/}\ \ Proofs of  Main Results \endhead
\medskip

\heading{Rational times}\endheading

The fundamental solution $\,E(t,x)=\sum_n\,\be({n^2t\over2}+nx)\,$ 
is intimately related to the Jacobi theta function
$$
\vartheta (\tau, x,y)=
\sum_n\,\be({\tau\over 2}(n-y)^2+nx-\frac 12 xy)\,.\tag 3.1
$$
Indeed, it is easy to see that
$$
\vartheta (\tau, x,y)=\be(-\frac 12 y(x-\tau y))\,E(\tau,x)\,.\tag 3.2
$$
Recall, that by the well-known transformation property of $\,\vartheta$, 
\cite{Eichler}, for any unimodular matrix 
$\,g=\pmatrix a & b\\
c & d\endpmatrix\in SL(2,\Bbb Z)$, we have
$$\multline
\vartheta (\tau, x,y)=\\ 
\varkappa(g) |c\tau+d|^{-\frac 12}
\be(\frac 14 \eta(ax+by)-\xi(cx+dy))\,
\vartheta ({a\tau+b\over c\tau+d},ax+by-\frac 12\xi, cx+dy-\frac 12\eta),
\endmultline
\tag 3.3
$$
where $\,\xi = ab\,\mod 2$, $\,\eta=cd\,\mod 2$, and $\,\varkappa(g)$ 
is an eighth root of $\,1$; $\,\varkappa(g)$ depends on the matrix 
$\,g$ and the choice of $\,\xi$ and $\,\eta$. 


Let $\,t$ be a rational number in $\,(0,2)$, $\,t= p/q$, 
$\,p\,$ and $\,q\,$ are co-prime. As it was explained in Section 1, 
we choose the finite continued fraction representation 
$\, p/q=[a_0,a_1,\dots,a_{n-1},a_n]\,$ with {\it even\/} $\,n$. 
If $\,{p_{n-1}/q_{n-1}}=[a_0,a_1,\dots,a_{n-1}]$, then 
$\,p q_{n-1}-q p_{n-1}=1$, so, the matrix 
$\,g=\pmatrix q & -p\\
q_{n-1} & -p_{n-1}\endpmatrix\,
$ 
is in $\,SL(2,\Bbb Z)$. Using (3.3) with this $\,g$, and (3.2), 
we obtain
$$\multline
E(t,x)=\vartheta (t,x,0)=\varkappa(g)\sqrt{q}
\be(\frac 14x(q\eta-q_{n-1}\xi))
\vartheta (0,qx-\frac 12\xi, q_{n-1}x-\frac 12\eta)=\\
\varkappa(g)\sqrt{q}
\be(\frac 14 (qx-\frac 12\xi)(q_{n-1}x-
\frac 12\eta)+\frac 14 x(q\eta-q_{n-1}\xi)) E(0, qx-\frac 12\xi)=\\ 
\hfill
\varkappa(g)\sqrt{q}
\be(-\frac 12q q_{n-1} x^2+\frac 12q\eta x-\frac 18\xi\eta)
\sum_n\delta(n-qx+\frac 12\xi).
\endmultline
$$
This proves part 1) of Theorem {\rom I\/}.


To prove part 2), we need an auxiliary result, which will 
be used in the proofs of other results as well.

\proclaim{Lemma 3.1} Let $\,\chi(\cdot)\,$ be a continuously 
differentiable, non-negative function on $\,\Bbb R_+\,$ with support 
in the interval $\,[1/2,\, 2]$. Then, first, 
$$
\sum\limits_{n=0}^{\infty}\chi(2^{-j}n)\,=\,
2^{j}\,\int_0^{\infty}\chi(y)\,dy\,
+\,O(1),\quad\text{as}\quad j\to\infty\,,\tag 3.4
$$
and second,
$$
\sum\limits_{n=2^{j-1}}^{2^{j+1}}
\big |\chi(2^{-j}(n+1)) - \chi(2^{-j}n)\big |\le \varkappa, \tag 3.5 
$$
for all $\,j\ge 0$, with some constant $\,\varkappa$. 
\endproclaim
\demo{Proof} The proof of (3.4) 
is by direct application of the well-known 
Euler-Maclaurin summation formula (see, e.g., \cite{Edwards, Section 6.2}),
$$
\sum\limits_{n=M}^{N} f(n) = \int_M^N f(y)\,dy + 
\frac 12\,\left( f(M)+f(N)\right) +
\int_M^N (y-[y]-\frac 12) f^{\prime}(y)\,dy,
$$
which is valid for any continuously 
differentiable function $\,f\,$ on the interval $\,[M,N]$.

To prove (3.5), apply the Newton-Leibniz formula.
\qed
\enddemo
\medskip
 
Now, with the help of Lemma 3.1 and Remark 1.2, 
part 2) of Theorem {\rom I} follows 
from Corollary 2.2, Corollary 2.4 and Theorem 2.5. 
\qed
\medskip
 
Let $\,t\in (0,2)\,$ be an irrational number and let 
$\,[a_0,\,a_1,\,a_2,\,a_3,\dots]\,$ be its continued fraction expansion, 
$$
t\,=\,a_0\,+\,
\cfrac1\\
a_1+\cfrac1\\
a_2+\cfrac1\\
\dots
\endcfrac
$$
The  integers $\,a_0,\,a_1,\,a_2,\dots\,$ can be found from the 
recurrent relations 
$$
a_{k+1}=\big[{1\over t_k}\big],\qquad t_{k+1}={1\over t_k} - a_{k+1}\,,\tag 3.6a
$$
and the initial conditions  
$$
a_0=\big[{ t}\big],\qquad t_0 = t-a_0\,.\tag 3.6b
$$
As usual, $\,\big[r\big]\,$ denotes the largest integer not exceeding $\,r$.

The finite parts $\,[a_0,\,\,a_1,\,a_2,\dots,\,a_n]$, $\,n=1,\,2,\dots$,  
of our infinite continued fraction, 
sum up to the rational numbers $\,p_n/q_n\,$ -- convergents of $\,t$. The numerators and denominators of the convergents 
can be found using the relations
$$
p_{k+1} = a_{k+1} p_k + p_{k-1},\qquad q_{k+1} = a_{k+1} q_k + q_{k-1},
\tag 3.7a
$$
and the initial conditions  
$$
p_{-1}=1,\quad q_{-1}=0, \quad p_0=a_0,\qquad q_0= 1.
\tag 3.7b
$$
Of course, $\,{p_n/ q_n}\to t$. Also, we have 
$$
{1\over q_n (q_n+q_{n+1})}\,<\,|\,t - {p_n\over q_n}\, |\,<\,
{1\over q_n q_{n+1}}<{1\over q_n^2}. \tag 3.8
$$
For our purposes, it is important to know how fast the denominators 
$\,q_n\,$ grow. The celebrated result of Khinchin and L\'evy 
(see \cite{Khinchin, 1936}, \cite{L\'evy, 1937}) answers 
this question in the following sense: {\it for almost all\/} $\,t$,
$$
{\ln q_n \over n}\,{\underset{n\to\infty}\to\to}\,\ln \rho_*, \tag 3.9a
$$
where $\,\rho_*\,$ is given by
$$
\ln\,\rho_*\,=\,{\pi^2\over 12\,\ln\,2}\,.\tag 3.9b
$$
This means, in particular, that for any $\,t\,$ from the set defined 
by the Khinchin-L\'evy theorem and for all sufficiently large $\,j$, 
there exists $\,{p_n\over q_n}$, the $\,n^{th}\,$ convergent to $\,t$, 
with $\,q_n = 2^j\,\cdot\,2^{j\epsilon_j}$, where 
$\,n = [j\log_{\rho_*} 2]\,$ and $\,\epsilon_j\to 0\,$ 
as $\,j\to\infty$. Hence, part (i) of Theorem {\rom{II}} follows 
from Remark 1.2 and Corollaries 2.2 and 2.4.
\medskip

The equations (3.7) show that irrationals with bounded 
quotients belong to $\,\Cal I(0)$. 
Thus, the second statement of Theorem {\rom{II}} 
will follow from Theorem {\rom{III}}. Since inequality (1.9) 
with $\,\sigma=0\,$ and $\,c_t=1\,$ holds true for every 
irrational $\,t$, as (3.7a) shows, 
the third statement of Theorem {\rom{II}} follows from part (ii) 
of Theorem {\rom{III}}. 
We, therefore, turn to the proof of Theorem {\rom{III}}.
\medskip
\proclaim{Proposition 3.2} If $\,t\in\Cal I(\le\sigma)\,$ 
for some $\,\sigma\ge 0$, then 
$$
E(t,\cdot)\in B^{{}^{-{1+\sigma\over 2+\sigma}}}_{\infty}\,\cap\,
\lceil B\rceil^{{}^{-{1+\sigma\over 2+\sigma}}}_{\infty}\,. \tag 3.10
$$
\endproclaim 
\demo{Proof} Recall, that $\,t\in\Cal I(\le\sigma)\,$ means that 
the denominators $\,q_n\,$ of the convergents to $\,t\,$ 
satisfy the inequality
$$
q_{n+1}\le C_t\,q_n^{1+\sigma}\,,\tag 3.11
$$
for all sufficiently large $\,n$. In order to prove (3.10), we 
have to obtain the appropriate 
estimates (1.5) for truncated exponential sums. We achieve this by applying 
Corollaries 2.2 and 2.4. The sums in (1.5a) and (1.5b) will be treated 
similarly, so we shall work only with the one that is involved in the 
definition of the space 
$\,\lceil B\rceil^{{}^{-{1+\sigma\over 2+\sigma}}}_{\infty}$. 
Our goal is to show that there is a constant $\,\lceil C\rceil$ 
such that, for all $\,x$, the inequality
$$
\big | \sum\limits_{m=2^{j-1}+1}^{2^{j+1}} \be ({m^2t\over 2}+mx)\big |
\le \lceil C\rceil\,2^{\alpha j},
\tag 3.12
$$
holds true for all sufficiently large $\,j$, where 
$$
\alpha={1+\sigma\over 2+\sigma }.\tag 3.13 
$$
By Legendre's theorem, 
$\,|t-(p_n/q_n)|\le 1/(2q_n^2)$, for any convergent $\,p_n/q_n$. 
Hence, Corollary 2.2 provides an estimate,
$$
\big | \sum\limits_{m=2^{j-1}+1}^{2^{j+1}} \be ({m^2t\over 2}+mx)\big |
\,\le\,\tilde C \,\left({2^j\over\sqrt{q_n}}\,+\,\sqrt{q_n}\right), 
\tag 3.14 
$$
for an arbitrary $\,n$. It remains to show that
$$
{2^j\over\sqrt{q_n}}\,+\,\sqrt{q_n}\,\le\, C\,2^{\alpha j},\tag 3.15a
$$
for every sufficiently large $\,j$, provided $\,n\,$ is chosen appropriately.

Define the exponents 
$\,s_n\,$ so that 
$$
2^{s_n}\,=\,q_n\, .\tag 3.16
$$
In order that (3.15a) be true, 
the following inequalities  ought to be satisfied: 
$\,\alpha \ge 1/2$ (which is the case when $\,\sigma\ge 0$), and 
$$
{s_n\over 2\alpha}\,\le\,j\,\le {s_n\over 2(1-\alpha)}\,.\tag 3.17a
$$
We would like to have (3.15a) with $\,q_n\,$ replaced by $\,q_{n+1}\,$ as well. 
To achieve this, we impose a stronger requirement, which uses 
the assumption (3.11).  Namely, we require that
$$
{2^j\over\sqrt{q_{n+1}}}\,+\,\big(C_t\,q_n^{1+\sigma}\big)^{1/2}\,
\le\, C\,2^{\alpha j},\tag 3.15b
$$ 
This inequality is satisfied provided 
$$
{s_n (1+\sigma)\over 2\alpha}\,\le\,j\,
\le {s_{n+1}\over 2(1-\alpha)}\,.\tag 3.17b
$$
Since $\,\alpha = {1+\sigma\over 2+\sigma}$, the right end of the interval 
(3.17a), $\,{s_n\over 2(1-\alpha)}$, coinsides 
with the left end of the interval (3.17b), $\,{s_n (1+\sigma)\over 2\alpha}$. 
Since the exponents $\,s_n\,$ grow to infinity, 
the above argument shows that the desired 
estimate (3.12) holds for all sufficiently large $\,j$. Proposition follows. 
\qed
\enddemo

\proclaim{Proposition 3.3} If $\,t\in\Cal I(\ge\sigma)\,$ 
for some $\,\sigma\ge 0$, then
$$
E(t,\cdot)\notin 
\left(
\cup_{\epsilon>0}B^{{}^{-{1+\sigma\over 2+\sigma}+\epsilon}}_{\infty}
\right)\,
\cup\,
\left(\cup_{\epsilon>0}\lceil 
B\rceil^{{}^{-{1+\sigma\over 2+\sigma}+\epsilon}}_{\infty}
\right)\,. 
$$
\endproclaim 
\demo{Proof} Let $\,t\in\Cal I(\ge\sigma)$. Recall, that this means 
that 
$$
q_{n+1}\ge c_t\,q_n^{1+\sigma} \tag 3.18
$$ 
for an infinite number of $\,n$'s. 
We are going  to prove that 
$\;E(t,\cdot)\notin\cup_{\ve>0} B^{{}^{-{\alpha}+\epsilon}}_{\infty}$, where 
$\,\alpha\,$ is defined in (3.13). It is sufficient to show that 
there exists an infinite number of  $\,j$'s, such that 
$$  
\sup_x\,
\big | \sum\limits_{m=2^{j-1}}^{2^{j+1}}
\chi(2^{-j}m) \be ({m^2t\over 2}+mx)+
\chi(2^{-j}m) \be ({m^2t\over 2}-mx)
\big |
\ge 
 C\,2^{\alpha j}\,. \tag 3.19
$$
We first note that if 
$\,p /q \,$ is such that 
$$
|t-{p \over q }|\,\le\,K\,2^{-2j}\,,\tag 3.20
$$
then, by Theorem 2.6, we can replace $\,t\,$ by $\,p /q \,$ 
in (3.19). After this replacement, we can use Theorem 2.5 to get  
the following estimate:
$$\multline
\sup_x\,
\big | \sum\limits_{m=2^{j-1}}^{2^{j+1}}
\chi(2^{-j}m) \be ({m^2 p \over 2q }+mx)+
\chi(2^{-j }m) \be ({m^2p \over 2q }-mx)
\big |\\
\hfill\ge\,\tilde c\,\min\,\{2^{j/2},\;{2^{j}\over\sqrt{q }}\}\,.
\hfill
\endmultline
\tag 3.21
$$
Note, that we have simplified the  right sides of  
inequalities  (2.8) by observing that 
$$
\sum\limits_{m=2^{j-1}}^{2^{j+1}} \omega_{m}+\omega_{-m}\,=\,
2\,\sum\limits_{m=2^{j-1}}^{2^{j+1}}
\chi(2^{-j}m)\, \ge\,\sqrt{2}\,\tilde c\;2^{j},
$$
where the last inequality is a consequence of Lemma 3.1. 
Now, (3.19)  will be proved if we show that 
$$
\min\,\{2^{j/2},\;{2^{j}\over\sqrt{q}}\}\,\ge\,
C\,2^{\alpha j}, \tag 3.22
$$
for an infinite number of $\,j$'s, provided $\,p /q \,$ 
are chosen appropriately. 

Our plan is to use the convergents $\,p_n/q_n\,$ 
in the above argument. We first  check (3.20). Since 
$\,|t-(p_n/q_n)|<1/(q_nq_{n+1})\,$ for all $\,n$, we have 
$$
|t-{p_n\over q_n}|\,<\,{1\over c_t q_n^{2+\sigma}}\,
=\,(1/c_t)\,2^{-s_n(2+\sigma)},\tag 3.23
$$ 
for those $\,n$, for 
which (3.18) takes place. Here and further on, $\,s_n=\log_2 q_n$. 
Thus, to satisfy (3.20),  it is sufficient to require that 
$$
j\,\le\,\frac 12\,s_n(2+\sigma)\,.\tag 3.24a
$$
Next, in order to have (3.22), it is sufficient to ask for 
the following inequality to hold:
$$
j\,\ge\,{s_n\over 2(1-\alpha)}-1\,=\,\frac 12\,s_n(2+\sigma)\,.\tag 3.24b
$$
The interval $\,[\frac 12\,s_n(2+\sigma)-1,\,\frac 12\,s_n(2+\sigma)]\,$ 
always contains an integer $\,j$, and, since $\,s_n\to\infty\,$ (we use 
only those $\,n\,$ for which (3.18) takes place), 
(3.19) is satisfied for an infinite number of $\,j$'s. 
This completes the proof of the proposition. 
\qed
\enddemo

\centerline{REFERENCES}
\vglue 1pc
 
\ref\by J. Bergh and J. L\"ofstr\"om
\book Interpolation spaces, an introduction
\publ Springer
\publaddr Berlin
\yr 1976
\endref

\ref\by M.V. Berry and J. Goldberg
\paper Renormalization of curlicues
\jour Nonlinearity
\vol 1
\pages 1-26
\yr 1988
\endref

\ref\by E. Bombieri 
\paper On Vinogradov's Mean Value Theorem 
and Weyl sums
\inbook  Proceedings of the conference on
"Automorphic forms and analytic number theory",
Montr\'eal, June 6-10, 1989
\pages 7-24
\publ CRM
\publaddr Montr\'eal 
\yr 1990
\endref

\ref\by J. Bourgain
\paper Fourier transform restriction phenomena
for certain lattice subsets and applications
to nonlinear evolution equations.{\rom I\/}. 
Schr\"odinger equations
\jour Geom. Funct. Anal.
\vol 3
\issue 2
\yr 1993
\endref

\ref\by E. A. Coutsias and N. D. Kazarinoff
\paper Disorder, renormalizability, theta functions 
and Cornu spirals
\jour Physica D
\vol 26
\pages 295-310
\yr 1987
\endref

\ref\by F. M. Dekking and M. Mend\`es France
\paper Uniform distribution modulo one: 
a geometric view point
\jour J. Reine Angew. Math.
\vol 329
\pages 143-153
\yr 1981
\endref

\ref\by J.-M. Deshouillers
\paper Geometric aspect of Weyl sums
\inbook Elementary and analytic theory of numbers, 
Banach Center publications, volume 17
\pages 75-82
\publ PWN-Polish Scientific Publishers
\publaddr Warsaw
\yr 1985
\endref

\ref\by H. M. Edwards
\book Riemann's Zeta Function
\publ Academic Press
\publaddr New York and London
\yr 1974
\endref

\ref\by M. Eichler
\book Introduction to the theory of algebraic numbers
and functions
\publ Academic Press
\publaddr New York and London
\yr 1966
\endref

\ref\by H. Fiedler, W. Jurkat and O. K\"orner
\paper  Asymptotic expansion of finite theta series
\jour Acta Arithmetica
\vol 32
\pages 129-146
\yr 1977
\endref

\ref\by J. H. Hannay and M. V. Berry
\paper Quantization of linear maps on a torus -- Fresnel 
diffraction by a periodic grating
\jour Physica D
\vol 1
\pages 267-290
\yr 1980
\endref

\ref\by G. H. Hardy and J. E. Littlewood
\paper Some problems of Diophantine approximation
\jour Acta Math
\vol 37
\yr 1914
\pages 155-239
\endref

\ref\by W. B. Jurkat and J. W. van Horn
\paper The proof of the central limit theorem for theta sums
\jour Duke Math. J.
\vol 48
\pages 873-885
\yr 1981
\endref

\ref\by W. B. Jurkat and J. W. van Horn
\paper  On the central limit theorem for theta series 
\jour Michigan Math. J.
\vol 29
\pages 65-77
\yr 1982
\endref

\ref\by L. Kapitanski and I. Rodnianski
\paper Regulated smoothing for Schr\"odinger evolution
\jour International Math. Research Notices
\vol 2
\pages 41-54
\yr 1996
\endref

\ref\by L. Kapitanski, I. Rodnianski and K. Yajima
\paper On the fundamental solution
 of a perturbed harmonic oscillator
\jour TMNA 
\vol 9
\issue 1
\yr 1997
\pages 77-106
\endref

\ref \by A. Ya. Khinchin
\paper  Zur metrischen Kettenbruchtheorie
\jour  Compositio Mathematica \yr 1936
\vol 3\issue 2 \pages 275-285
\endref

\ref \by A. Ya. Khinchin 
\book Continued fractions
\bookinfo Translated from the 3rd Russian edition of 1961
\publ The University of Chicago Press 
\yr 1964
\endref

\ref \by P. L\'evy
\book Th\'eorie de l'addition des variables al\'eatoires
\publ  
\publaddr Paris
\yr 1937
\endref

\ref\by P. I. Lizorkin
\paper On bases and multipliers for the spaces $\,B^r_{p,\Theta}(\Pi)$
\jour Trudy Mat. Inst. Steklov
\vol 143
\pages 88-104
\yr 1977
\endref

\ref\by J. Marklof
\paper Limit theorems for theta sums  
\paperinfo see this volume
\moreref 
\book Limit theorems for theta sums 
with applications in quantum mechanics 
\bookinfo Dissertation
\publaddr Ulm
\yr 1997
\endref

\ref \by H. L. Montgomery
\book Ten lectures on the interface between analytic number theory 
and harmonic analysis
\bookinfo CBMS, Regional Conference Series in Mathematics, Number 84
\publ American Mathematical Society
\publaddr Providence, Rhode Island
\yr 1994
\endref

\ref\by L.J. Mordell 
\paper The approximate functional formula
for the theta function
\jour J. London Math. Society
\vol 1
\pages 68-72
\yr 1926
\endref

\ref\by K. I. Oskolkov
\paper A class of I. M. Vinogradov's series and its applications 
in harmonic analysis
\inbook Progress in Approximation Theory, an International perspective
\eds A. A. Gonchar and E. B. Staff
\paperinfo Springer Series in Computational Mathematics, 19
\publ Springer-Verlag
\publaddr New York
\yr 1992
\endref

\ref\by P. Sarnak
\paper Asymptotic behavior of periodic orbits of the 
horocycle flow and Eisenstein series
\jour Comm. Pure Appl. Math.
\vol 34
\yr 1981
\pages 719-739
\endref

\ref\by P. Sarnak
\paper Class numbers of indefinite binary quadratic forms
\jour J. Number Theory
\vol 15
\yr 1982
\pages 229-247
\endref

\ref\by Schmidt
\book Diophantine approximation
\bookinfo Lecture Notes  in Mathematics, 785
\publ Springer-Verlag
\publaddr Berlin Heidelberg New York
\yr 1980
\endref

\ref \by H. Triebel
\book Theory of function spaces
\publ Birkh\"auser
\publaddr Basel
\yr 1983
\endref

\ref\by K. Yajima
\paper Smoothness and non-smoothness of the fundamental solution 
of time dependent Schr\"odinger equations
\jour Comm. Math. Phys.
\vol 181
\issue 3
\pages 605-629  
\yr 1996
\endref

\end